\documentclass[prc,showpacs,nofootinbib]{revtex4} 
\usepackage{rotating}
\usepackage{epsfig}
\usepackage{dcolumn}
\usepackage{graphicx}
\begin{document}

\title{La-138/139 isotopic data and neutron fluences for Oklo RZ10 reactor}

\author{C. R.  Gould}
\affiliation{Physics Department, North Carolina State University, 
Raleigh, NC 27695-8202, USA}
\affiliation{Triangle Universities Nuclear Laboratory, Durham, 
NC 27708-0308, USA}
\email{chris_gould@ncsu.edu}
\author{E. I.  Sharapov}
\affiliation{Joint Institute for Nuclear Research, 141980 Dubna, Moscow region, Russia}

\vspace{2cm}

\date{July 22, 2012}

\begin{abstract}

\noindent {\bf Background:} 
Recent years have seen a renewed interest in the Oklo phenomenon,
particularly in relation to the study of 
time variation of the fine structure constant $\alpha$.  The neutron fluence is one of
 the crucial parameters for Oklo reactors. Several approaches to its determination 
were elaborated in the past.
\\
{\bf Purpose:} We consider whether it possible to use 
the present isotopic $^{138}$La$-$$^{139}$La data for RZ10 as an additional 
indicator of neutron fluences in the active cores of the reactors.
\\
{\bf Results:} We calculate the dependence of the Oklo   $^{138}$La abundance  on  neutron fluence and elemental Lanthanum  concentration. \\
{\bf Conclusion:} The neutron fluence in RZ10 can be deduced from
Lanthanum isotopic data, but requires reliable data on the primordial elemental abundance. Conversely, if the fluence is known,
the isotope ratio provides information on the primordial Lanthanum abundance that is not otherwise easily determined. 
\end{abstract}
\vspace{1pc}
\maketitle

The Oklo natural nuclear reactors \cite{Nau91} continue
to excite scientific interest. Recent studies have been related to the implications of Oklo on the
problem  of time variation of the fine structure constant $\alpha$
\cite{Chris06,Petr06,Oneg010,Gould012}, and to the geological context
and operational issues of the reactors \cite{Salah011}. Zone RZ10 has
emerged as one of the most important because it was deep underground and apparently undisturbed over the 2 GY since the reactor ceased operating.
The neutron fluence in RZ10 is one of
 the crucial parameters, central to all analyses that seek to set bounds on the change in $\alpha$.   Several approaches to its determination 
were elaborated in the past. They rely primarily on changes in the Oklo isotopic abundances of
Nd due to neutron fission and capture reactions \cite{Nau91}. 
In this short note we argue that the Oklo isotopic abundances of
Lanthanum can provide an
independent confirmation of the fluence if reliable elemental data on primordial
Lanthanum in Oklo are available. And if the fluence is known from other sources, the isotopic ratios provide information on the primordial Lanthanum concentration that is otherwise not easily determined.
\\

The terrestrial abundances for the   Lanthanum isotopes are 0.088 \% for
$^{138}$La and 99.91 \% for $^{138}$La 
\cite{NNDC}. The abundance of the extremely rare
 odd-odd isotope $^{138}$La (half-life $1.02 \cdot 10^{11}$ yr) is of a great interest
itself and is under study in neutrino-nucleosynthesis models
\cite{Bey07}.  However in the Oklo core RZ10 the $^{138}$La was found
to be present even in a lesser proportion, the   abundances being  0.0129 \% in the two samples SF84-1469 and 1480
\cite{Hida98}. This is
evidently due to a large cumulative  production of $^{139}$La in
fission of $^{235}$U by thermal neutrons (cumulative fission yield $Y_{235,139} = 0.063$), while $^{138}$La
production is negligible  ($Y_{235,138} = 3 \cdot 10^{-7}$), because this isotope is 
shielded by a stable precursors in the beta-decay chains of fission products.\\

For our case, we describe the time evolution (during the Oklo
phenomenon) of the number densities $N_A(t)$ of the  
isotopes of interest - $^{235}$U,  and $^{139}$La, by a simple set of
coupled differential equations involving effective cross sections and fluxes: 
\begin{eqnarray}
{dN_{235}\over dt}&=&-\hat\sigma_{235,tot}N_{235} \hat\phi +C \hat\sigma_{235,tot}N_{235}\hat\phi\\
{dN_{139}\over
dt}&=&-\hat\sigma_{139}N_{139}\hat\phi+\hat\sigma_{235,f}Y_{235,139}N_{235}\hat\phi.
\end{eqnarray}
In these equations, the subscripts 235, and 139 refer to $^{235}$U, 
and $^{139}$La, respectively, $tot$ and $f$ refer to the total
(absorption) and fission cross sections, and $C$ is the restitution
factor taking into account the production of $^{235}$U from the
radioactive decay of $^{239}$Pu which is produced after
neutron capture in  $^{238}$U.

As is standard, we take the average neutron flux, $\hat\phi$, to be constant. $^{138}$La is not produced in fission, so its time evolution at the time $t_1$ of shutdown of the reactor is given simply by $ N_{138}(t_1)=N_{138}(0)\exp(-\hat\sigma_{138}\hat\phi t_1)$. The solution for $N_{139}$ 
is (see also  Ref. \cite{Hid88b} for the analogous case of $^{157}$Gd):
\begin{equation}
N_{139}(t_1)=N_{139}(0)\exp(-\hat\sigma_{139}\hat\phi t_1) + N_{235}(0)Y_{235,139}\frac{\hat\sigma_{235,f}}
{(1-C)\hat\sigma_{235,tot} - \hat\sigma_{139}}[\exp(-\hat\sigma_{139}\hat\phi t_1)-
\exp(-(1-C)\hat\sigma_{235,tot}\hat\phi t_1) ].
\end{equation}

The cross sections are well known, and the starting $^{235}$U
elemental concentration is reliably determined from the remaining U in
the cores. We can therefore calculate the isotopic fraction (we call it the Oklo $^{138}$La abundance)
\begin{equation}
f_{138}(t_1)=N_{138}(t_1)/(N_{138}(t_1)+ N_{139}(t_1))  
\end{equation}
either as a function of the fluence, $\hat\phi t_1$ (if  $N_{139}(0)$
is known), or as a function of the primordial concentration $N_{139}(0)$ (if the fluence is
known). 

Our basic parameters are as follows: Primordial isotopic abundances (corrected for decay) 
$f_{138}(0)=0.090$\% and $f_{139}(0)=99.91$\%. U effective neutron 
cross sections \cite{Nau91} $\hat\sigma_{235,f}= 549$ b,
$\hat\sigma_{235,tot}= 656$ b, effective La cross sections  $ \hat\sigma_{139} =9$ b, and $ \hat\sigma_{138} =58$ b from \cite{NNDC}, 
  fission yield $Y_{235,139}=0.063$ \cite{Cro77}, and restitution
  factor $C=0.38$ \cite{Hida98}.  For the U-235 number density we take
  the value from our realistic model of the reactor RZ10
  (Ref. \cite{Chris06}, Table III) $N_{235}(0)= 0.068\times 10^{21}$
  cm$^{-3}$, with a total density of 3 g cm$^{-3}$ for the active core
  material. \\

We estimate the starting Lanthanum elemental concentrations using the procedure outlined by Hidaka et al. \cite{Hid88c}.  Defining the ratio today of primordial Lanthanum to total Lanthanum by $x$, and noting that $^{138}$La is all primordial, we have:

\begin{equation}
f_{138}(t_1)/f_{139}(t_1)=(0.090x + 0(1-x))/(99.91x +100(1-x)).
\end{equation}

The value measured today for the two RZ10 samples is 0.0129 \%. Again correcting for the small fraction of $^{138}$La that has decayed since the reactor ceased operating, we find the primordial fraction of Lanthanum in the two RZ10 samples to be $x = 0.145$, giving   
$N_{139}(0)$ = 14.4 ppm for SF84-1469 (concentration today = 99.6 ppm) and 10.1 ppm for SF84-1480 (concentration today = 70.0 ppm). \\

Plots of $ f_{138}(t_1)$ as a function of the fluence for these two starting concentrations are shown in Fig. 1. The intersections with the measured value of $ f_{138}(t_1)$ lead to fluence estimates of 0.34 and 0.49 kb$^{-1}$ respectively for the two samples, or an average fluence for RZ10 from Lanthanum  of $ 0.41 \pm 0.10 $ kb$^{-1}$. 
In our earlier work on RZ10 \cite{Chris06}, we used a
meta-sample fluence, an average of four values
determined from the standard Nd-isotopic concentration method. Individual values are of course expected to vary due to the heterogeneous nature of the reactor zone. Our meta-sample value was $0.65 \pm 0.19$ kb$^{-1}$. The data of Hidaka and Holliger \cite{Hida98} lead to a meta-sample value of  $0.63 \pm 0.12$ kb$^{-1}$. Our Lanthanum result, while lower, is therefore  seen to be not incompatible with these more reliable Nd values.  \\

\begin{figure}[htbp] 
  \centering
\includegraphics[width=5.67in,height=3.77in,keepaspectratio]{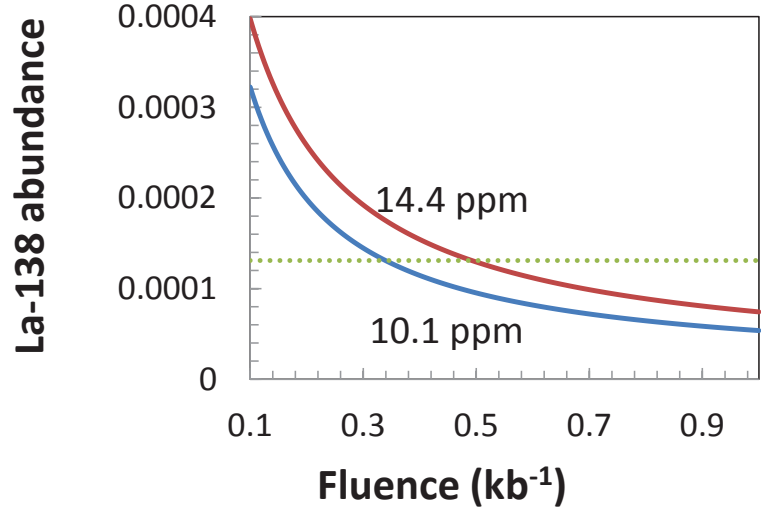}
\caption{ (Color online). The $^{138}$La abundance versus neutron fluence for Oklo active core RZ10. 
The curves are calculated according to Eq. (4) with primordial   
concentrations of Lanthanum equal to 10.1 and 14.4 ppm (see text). The intersections with the measured value of $ f_{138}(t_1)$ lead an average fluence for RZ10 of $ 0.41 \pm 0.10 $ kb$^{-1}$.
}
  \label{Fig1}
\end{figure}

Considerable variation in elemental concentrations from place to place is expected, in contrast to isotopic abundance data which tend to show less fluctuation and can be more reliably averaged inside or outside the reactor core.   
We therefore follow the alternate procedure, take the meta-sample fluence as a given at $0.65 \pm 0.19$ kb$^{-1}$, and use it to estimate
the starting Lanthanum elemental concentration in the core. We can compare this concentration to the two values determined earlier, and to measured values from outside the RZ10 core. There are three such values, and they vary widely: two measured about 2 m below the RZ10
core: SF84-1640, 1.8 ppm and SF84-1700, 13.5 ppm and one
measured about 0.5 m above the core, SF84-1400, a surprisingly high 199 ppm \cite{Hida98}.
The dependence of $f_{138}$ on $N_{139}(0)$ is shown in Fig. 2. 
The intersections with the measured value of $ f_{138}(t_1)$ lead to a concentration of $18.6 \pm 4.9$ ppm,  similar to SF84-1700, and close to the two concentrations determined from the isotopic ratio data itself.  \\

\begin{figure}[htbp] 
  \centering
 \includegraphics[width=5.67in,height=3.77in,keepaspectratio, angle = 
0 ]{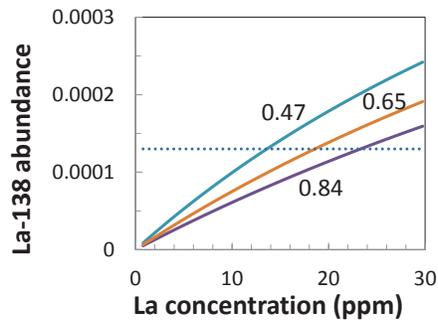}
\caption{(Color online). The $^{138}$La abundance versus primordial 
elemental Lanthanum abundance for Oklo active core RZ10. 
The curves are calculated according to Eq. (4) with a fluence of $0.655 
\pm 0.187$ kb$^{-1}$, leading to a predicted primordial abundance of $18.6 
\pm 4.9$ ppm .}
  \label{Fig2}
\end{figure}

Clearly many more samples within and outside the reactor zones would
be needed  to establish reliable baseline values, and to estimate uncertainties. 
Nevertheless, given the known variability of Oklo rare-earth elemental concentrations
we find this is an encouraging result. Determining starting elemental concentrations in the Oklo cores may be of value since there is evidence that 
the Oklo phenomenon is  potentially complicated by hints of
selective retention of rare earth elements depending on whether they
are of fissiogenic or non-fissiogenic origin \cite{Hida00}. There are also
indications that uraninite in Oklo incorporated lower 
amount of rare-earth elements during primordial mineralization \cite{Hida92}.  
Our present result for Lanthanum concentration may therefore be of
interest with respect to the study of uraninite formation processes.

\acknowledgments

We thank a referee for valuable suggestions on determining primordial elemental abundances.  This work was supported by the US Department of Energy,
Office of Nuclear Physics, under Grant No.
DE-FG02-97ER41041 (NC State University).

\end{document}